\documentclass[10pt,journal,compsoc]{IEEEtran}
\usepackage{graphicx}
\usepackage{paralist}
\usepackage{amsmath}
\usepackage{amsthm}
\usepackage{booktabs}
\usepackage{mathtools}
\usepackage[caption=false]{subfig}

\DeclarePairedDelimiter\floor{\lfloor}{\rfloor}
\usepackage[multiple]{footmisc}
\usepackage{xcolor}
\usepackage[linesnumbered,ruled,vlined]{algorithm2e}
\usepackage{soul}
\usepackage[normalem]{ulem}
\usepackage{amsmath}
\newcommand{\stkout}[1]{\ifmmode\text{\sout{\ensuremath{#1}}}\else\sout{#1}\fi}
\usepackage[bookmarks=false]{hyperref} 

\SetCommentSty{mycommfont}

\SetKwInput{KwInput}{Input}                
\SetKwInput{KwOutput}{Output}              

\newcommand{\myparagraph}[1]{\vspace{0.1cm}\noindent{\bf #1.}}

\ifCLASSOPTIONcompsoc
  \usepackage[nocompress]{cite}
\else
  \usepackage{cite}
\fi

\renewcommand{\tablename}{Table}

\renewcommand{\figurename}{Figure}

\ifCLASSINFOpdf
\else
\fi
\begin{document}
\title{Always on Voting: A Framework for Repetitive Voting on the Blockchain}

\author{Sarad Venugopalan\IEEEmembership{},
         Ivana Stan\v{c}\'{i}kov\'{a}, Ivan Homoliak\IEEEmembership{}

\IEEEcompsocitemizethanks{\IEEEcompsocthanksitem \textbf{Sarad Venugopalan currently has no affiliation. Ivana Stan\v{c}\'{i}kov\'{a} and Ivan Homoliak  are with Brno University of Technology, Czech Republic. \textbf{This is a pre-print of the  article accepted for publication in  IEEE Transactions on Emerging Topics in Computing. Digital Object Identifier \href{https://doi.org/10.1109/TETC.2023.3315748}{10.1109/TETC.2023.3315748}. IEEE Xplore link: https://ieeexplore.ieee.org/document/10260281/}} \protect\\
}
\thanks{}}

\markboth{}%
{Shell \MakeLowercase{\textit{et al.}}: Bare Demo of IEEEtran.cls for Computer Society Journals}
\IEEEtitleabstractindextext{%
\begin{abstract}
Elections repeat commonly after a fixed time interval, ranging from months to years.
This results in limitations on governance since elected candidates or policies are difficult to remove before the next elections, if needed, and allowed by the corresponding law. Participants may decide (through a public deliberation) to change their choices but have no opportunity to vote for these choices before the next elections.
Another issue is the peak-end effect, where the judgment of voters is based on how they felt a short time before the elections. 
To address these issues, we propose Always on Voting (AoV) -- a repetitive  voting framework that allows participants to
vote and change elected candidates or policies without waiting for the next elections.
Participants are permitted to privately change their vote at any point in time, while the effect of their change is manifested at the end of each epoch, whose duration is shorter than the time between two main elections. 
To thwart the problem of peak-end effect in epochs, the ends of epochs are randomized and made unpredictable, while preserved within soft bounds. 
 These goals are achieved using the synergy between a Bitcoin puzzle oracle, verifiable delay function, and smart contracts. 
\end{abstract}

}

\IEEEoverridecommandlockouts
\IEEEpubid{\makebox[\columnwidth]{2168-6750 © 2023 IEEE.  \hfill}
	\hspace{\columnsep}\makebox[\columnwidth]{ }}
\maketitle
\IEEEpubidadjcol
 \pagenumbering{gobble}

\IEEEdisplaynontitleabstractindextext

%
\IEEEpeerreviewmaketitle

\section{Introduction}
\label{sec:introduction}

Voting is an integral part of democratic governance, where eligible participants can cast a vote for their representative (candidate or policy) through a secret ballot.
The outcome is an announcement of winners through a tally of votes. 
In practice, the time interval between two regularly scheduled elections is usually large -- ranging from months to years.
Over time, a previously popular winning candidate (or policy) may have fallen out of favor with the majority of participants.
Therefore, we argue that a common lacking attribute in governance is the ability of participants to reverse or correct the previous decisions that were collectively voted for when new information is available after the election.
Reasons for poor decision making (i.e., error premise~\cite{Blum2016}) may arise from insufficient or false information, search engine manipulation~\cite{EpsteinRLW17}, social media manipulation~\cite{Goldzweig2019}, or from agenda setters~\cite{Borgesius2018}.
To deal with this issue, we propose a repetitive voting strategy, which gives its participants the ability to change their vote anytime they decide.
Nevertheless, even in a repetitive voting with fixed time intervals, participants remain exposed to constant manipulation attacks.
However, in contrast to standard voting with long time intervals, participants of repetitive voting might hold any elected candidate accountable by changing their vote choice.

A second concern is a peak-end effect, whose discovery in behavioral science is attributed to Nobel laureate Kahneman and his research collaborators~\cite{Kahneman1993}. Their study on the correlation of pain perception
over time indicated that duration plays a minor part in retrospective evaluations of aversive experiences.
The experiences are also dominated by discomfort at the worst and the final moments of episodes.
Carmon and Kahneman~\cite{Carmonphdthesis} found that how participants felt at the final moment of the experience was a good predictor of their overall experience evaluation responses.

There are many studies in political science analyzing and confirming the existence and impact of the peak-end effect that might be caused by
economic growth in the elections and pre-elections year~\cite{Dash2018,Healy2014,Wlezien2015}, increased spending on highly visible areas~\cite{QingyuanLi2019,Olejnik2021,Guinjoan2020},
private and government credit easing~\cite{Kern2021},
strategically planned welfare reforms~\cite{Wenzelburger2020},
and cash transfer in exchange for using school and health services by poor households~\cite{Galiani2019} (see  	background  on peak-end-effects in \autoref{ssec:Peaknew}).

\myparagraph{Our Approach}
We propose Always-on-Voting (AoV) that supports 1-out-of-$k$ candidate voting and runs on a blockchain\footnote{A Byzantine Fault Tolerant state machine replication protocol.}.
AoV has three key features: (1) it works in repetitive epochs, (2) voters are allowed to change their vote anytime before the end of each epoch (when the tally is computed), and (3) ends of epochs are randomized and unpredictable.
Only the supermajority of votes can change the previous winning vote choice at the end of each epoch.

In AoV, to thwart\footnote{The voting start and end times need to be known in advance to
	put in place the policies required to boost the elections campaign and
	entice a large number of voters over a short period of time. However,
	with AoV, though it is possible to make pre-elections promises, they run
	hollow after a while, and the spending budget is distributed over a
	longer period of time because of the repeated nature of voting and its
	uncertain tally timing, reducing the peak ‘awe’ effect, making it a less
	effective strategy to win votes.}
peak-end effects and decrease manipulation of participants, we randomize the time intervals between epochs of elections using public randomness and secure it with a verifiable delay function (VDF).
The tally time is random and unpredictable, precluding the interested parties from timely peak-end effect manipulations. One of the obstacles to implementing repetitive voting is increased resource expenditure (e.g., time \& money).  
To alleviate cost concerns, improve decentralization and to enhance security properties such as tamper resistance, we recommend our voting framework to be run on a public permissioned blockchain (see \autoref{ssec:permissioned}).

Several e-voting and blockchain voting solutions under various security requirements are compared in \autoref{ssec:blockchainevotingrelatedwork}.
In this work, we focus on addressing security challenges arising from introducing repetitive strategy, i.e., randomized tally times and repetitive voting.

\myparagraph{Contributions}
We make the following contributions.

\begin{compactenum}[i)]
	\item We identify two shortcomings in present governance systems for voting:
	a) the inability of participants to change their vote between two consecutive elections (e.g., that might be many months or a few years apart), and
	b) a manipulation of participants via peak-end effect (see \autoref{ssec:Peaknew}).
	
	\item We propose Always-on-Voting (AoV) framework for repetitive voting, which incorporates voting epochs and alleviates the shortcomings of present governance systems (see \autoref{sec:framework}).
	
	\item We propose the use of public randomness to determine when the current interval of voting should end using commitments to a future event in order
	to thwart the peak-end effect (see \autoref{ssec:vrf}).

	\item
	Finally, we analyze the randomness of the Bitcoin Proof-of-Work puzzle solution (hereafter referred to as a \textit{nonce}) and AoV entropy requirements in \autoref{ssec:rndnonce}.
\end{compactenum}

\section{Background}
\label{sec:Background}
In this section, we describe the preliminaries required to describe  our approach.

\begin{figure*}
	\centering
	\includegraphics[width=1.0\textwidth]{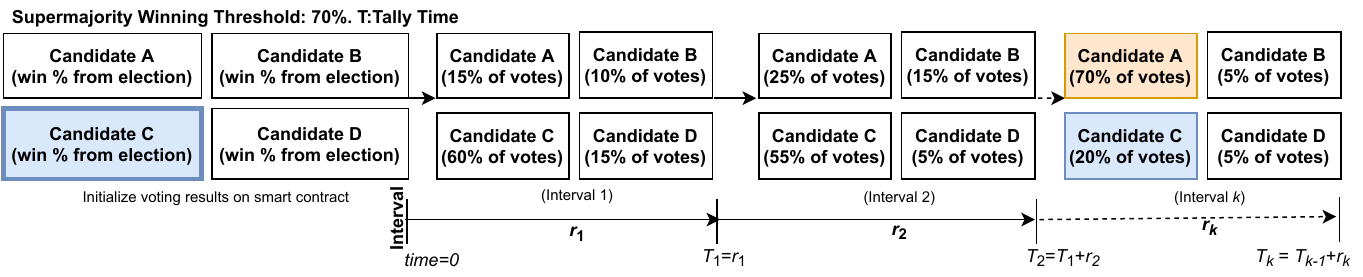}
	\vspace{-0.8cm}
	\caption{The time between two regular elections is divided into the fixed number $ft$
		of intervals (a.k.a., epochs). First, the ratios of votes for all vote choices (i.e., candidates) are initialized from the last election. Next, repeated voting within $k$ epochs results in a winning vote choice transition (from C to A). The new winner A is declared when she obtains a supermajority of total votes (i.e., 70\%) at interval $k$; $k\leq ft$ (see \autoref{sec:framework}).
		Note that $r_1,\ldots,~r_k$ are  randomized times that determine the length of the intervals $1,...,k$. The tally is computed at the end of each interval.}
	\label{fig:transition-epoch}
	\vspace{-0.3cm}
\end{figure*}

\subsection{The Blockchain}
\label{ssec:blockchains}
The data structure used in the blockchain represents an append-only
distributed ledger.
Its entries consist of transactions aggregated within ordered blocks.
The order of the blocks is agreed upon by a mutually untrusted honest majority of participants running a consensus protocol, i.e., consensus nodes (a.k.a., miners).
The blockchain is resistant to tampering by design since blocks are linked using a cryptographic hash function.
The blocks are considered irreversible (with overwhelming probability) after elapsing a time to finality.
Some blockchains are equipped with smart contract platforms that enable the users to write application code (i.e., smart contracts) and execute it.
All transactions sent to the blockchain are executed and validated by mutually untrusted consensus nodes.
In this way, smart contracts enable trusted code execution, where the trust relies on the honest threshold of consensus nodes (e.g., greater than 50\% in Proof-of-Work protocols and 67\% in Byzantine Fault Tolerant protocols).

\subsection{Bitcoin Proof-of-Work Puzzle}
\label{ssec:puzzlepow}
Bitcoin uses
Proof-of-Work (PoW) to achieve consensus among consensus (miner) nodes.
A block in Bitcoin is generated once every 10 minutes, on average.
A block consists of  2 parts, the header and body.
The Bitcoin header has a field called \textit{nBits} that encodes Bitcoin mining difficulty.
The \textit{merkle root} field stores the root of the Merkle hash tree corresponding to the transactions in the block.
The Bitcoin PoW puzzle is a lottery solved by finding a nonce $s$;  such that the SHA-256 hash of the Bitcoin block header that contains $s$ is lower than $target$.\footnote{See \url{https://learnmeabitcoin.com/technical/target}.}
The 32-bit nonce itself is a part of the header and adjusted using a random trial-and-error approach until a solution is found.
The mining difficulty changes every 2016 blocks (i.e., $\sim$every 2 weeks):
it is decreased if it took more time to mine 2016 blocks and increased if less time was required.
In this work, we use Bitcoin headers from already generated blocks as the source of randomness (see details in \autoref{ssec:rndnonce}).

\subsection{Verifiable Delay Function}
\label{ssec:VDF}
The functionality of VDF~\cite{Boneh2018} is similar to a time lock,\footnote{Time locks are computational problems that can only be solved by running a continuous computation for a given amount of time.} but in addition to it, by providing a short proof, a verifier may easily check if the prover knows the output of the VDF.
The function is effectively serialized, and parallel processing does not help to speed up VDF computation.
A moderate amount of sequential computation is required to compute VDF.
Given a time delay $t$, a VDF  must satisfy the following conditions:
for any input $x$, anyone equipped with commercial hardware can find $y$ = VDF($x, t$) in $t$ sequential steps, but an adversary with $p$ parallel processing units must not distinguish $y$ from a random number
in significantly fewer steps.  
For our purposes, the value of $t$ is fixed once it is determined. Therefore, to simplify our notation, we use VDF($x$) instead of VDF($x,t$) in the remaining text.
Further, given output $y$ of VDF, the prover can supply a proof $\pi$ to
a verifier, who may check the output $y ~=~ \text{VDF}(x)$ using $\pi$ in logarithmic time w.r.t. time delay $t$ (i.e., $VDF\_Verify (y,\pi) \stackrel{?}{=} True$).

Finally, the safety factor $A_{max}$ is defined as the time ratio that the adversary is estimated to run VDF computation faster on proprietary hardware as opposed to a benign VDF computation using commercial hardware (see Drake~\cite{Drake2018}).
CPU overclocking records~\cite{SAMUEL2020} indicate that $A_{max}=10$ is a reasonable estimate.

\subsection{E-Voting}
\label{ssec:votingprotocols}

Typically,  e-voting approaches have the setup phase, the registration phase,
followed by the voting and tally phases.
E-voting approaches include actors such as the election authority, candidates, and participants (i.e., voters).
In recent years, many blockchain-based e-voting approaches emerged (e.g.,~\cite{McCorrySH17,DBLP:conf/fc/SeifelnasrGY20,yu2018platform,killerprovotum,venugopalan2021bbbvoting}) since blockchains enable not only to instantiate the immutable public bulletin board required for e-voting~\cite{Kiayias2002} but also provide other features such as censorship-resistance and correct execution of smart contract code, which are beneficial in this context, e.g., to verify the correctness of submitted (encrypted/blinded) votes and compute the tally in a publicly verifiable fashion.
Also, blockchains contribute to end-to-end verifiability~\cite{benaloh2015end,killerprovotum} as well as universal verifiability~\cite{Kiayias2002,yu2018platform}.

\subsection{Peak End Effects}
\label{ssec:Peaknew}
In political science, the consequences of peak-end effects on voting have been extensively reviewed.
We summarize the results of such studies in the following.

Healy and Lenz~\cite{Healy2014} studied the bias in voter response to the elections-year economy. It showed that respondents put 75\% weight on the elections-year and 25\% weight on the year before, whereas the first 2 years of 4 years considered were insignificant. They showed that when the respondents were presented with easy-to-follow information on economic growth for all 4 years, the bias significantly decreased.
Wlezien~\cite{Wlezien2015} showed that voters weighed their decision not only on the elections-year economic indicators but also on the pre-elections year. The author attaches equal importance to the pre-elections year in influencing voter decisions.
Dash and Ferris~\cite{Dash2018} found the impacts of income growth during the election year as more significant in deciding the incumbents' re-election chances.

Kern and Amri~\cite{Kern2021} demonstrated that both private and government credit easing was common during the election year to court votes and stimulate credit growth.
Li et al.~\cite{QingyuanLi2019} showed the existence of a pattern to political investment cycles -- significantly more spending occurred during the election year.
Olejnik~\cite{Olejnik2021} conducted a study on Polish local government investment expenditure during the election year, indicating a significant increase in highly visible areas such as public infrastructure, tourism, and culture.
Guinjoan and Rodon~\cite{Guinjoan2020} observed a similar pattern of increased spending in highly visible areas such as local festivities during the election year.
Voters without expert knowledge in evaluating the investments may base their vote on the visibility of the actions taken.
Wenzelburger et al.~\cite{Wenzelburger2020} presented empirical evidence on how governments strategically plan welfare reforms to expand their benefits as the elections approach.
Galiani et al.~\cite{Galiani2019} indicated the peak-end pattern of the experiment on cash transfer in exchange for using school and health services by poor households in Honduras.
The study showed the increased vote share of the incumbent party in the elections, indicating sensitivity to recent economic activity.

Aguiar‐Conraria et al.~\cite{Conraria2019} examined the role of government transparency, i.e., the disclosure of relevant information. For the municipalities whose transparency indexes were high, voters rewarded policies that brought long-term benefits.   
The results of Carlin et al.~\cite{Carlin2021} showed that restrictions on relevant information distorted the ability of voters to choose in their best interests. On the other hand, transparent governance and independent media allowed voters to hold them accountable.

\section{System \& Adversary Model}
\label{sec:model}

\subsection{System Model}
\label{ssec:systemmodel}
Our model has the following main actors and components:
$\romannumeral
1$) A \textit{participant} ($P$) who partakes in governance by casting a vote for her choice or candidate.
$\romannumeral
2$) \textit{Election Authority} (EA) is responsible for validating the eligibility of participants to vote in elections, registering them, and shifting between the phases of the voting. A single EA  might be replaced by multiple election authorities (EAs) to improve decentralization. For example, a quorum of $>2/3^{rd}$ of its EAs must be in agreement to make election decisions. In our case, a single election authority is considered for simplicity.
$\romannumeral
3$)
A \textit{smart contract} (SC) collects the votes, acts as a verifier of valid voting, enforces the rules of the election  and verifies the tallies of votes.
$\romannumeral
4$)
\textit{Bitcoin Puzzle Oracle} (BPO) provides an off-chain data feed from the Bitcoin network and supplies the requested Bitcoin block header (BH) when it is available on the Bitcoin network.
$\romannumeral
5$)
A \textit{VDF prover} is any benign party in the voting ecosystem who computes the output of VDF and supplies proof of its correctness to SC.

\subsection{Adversary Model}
\label{ssec:advmodel}
The adversary in our voting framework with respect to epoch triggering is a Bitcoin mining adversary $Adv_{min}$.
This adversary is static and has bounded computing power, i.e., it is unable to break used cryptographic primitives under the standard security assumptions.
$Adv_{min}$ can mine on the Bitcoin blockchain.
Her goal is to find a solution to the Bitcoin PoW puzzle that also triggers the end of the current voting interval, thereby influencing the end time of epoch.
Our voting framework uses a function of the Bitcoin block header (BH) inclusive of its  PoW solution $s$, i.e., $f(BH)$ to trigger the end of the current voting interval.
Such a  manipulation would potentially enable $Adv_{min}$ to prematurely finish the current interval and start the next one (see Section~\ref{ssec:vrf}).

Finally, we assume that $EA$ verifies identities honestly and supply addresses of only verified participants to $SC$.

\section{Solution outline \& Design Goals }
\label{sec:solutionoutlinegoals}

\subsection{Solution Outline}
\label{ssec:outline}

In certain forms of governance (such as with traditional governance), elections are  repeated only after many months/years. Here, its participants will have to wait until the next election to change their vote. Due to the fixed time allocated for governance to the winning candidate, it is not possible to change this elected candidate before the allocated time for governance expires. During the time period between 2 consecutive elections, the incumbent may have fallen out of favor to a majority of the participants but current voting frameworks have no option to re-vote until the time to next election has elapsed. This results in limitations on governance since elected candidates or policies are difficult to remove before the next elections. This is an issue we address in our voting framework by breaking it down into smaller repeated epochs (along with thwarting the peak-end-effects in elections).

Our solution presents a 1-out-of-$k$ voting framework repeated over time.
	When the voting is repeated over a fixed time epoch, its working is similar to voting  carried out in a traditional election.
	However, in our proposed framework,  we introduce three main changes. (1) The time interval between consecutive elections is shorter when compared to regularly held elections (repeated after months or years). (2) A single trigger is used to end the current voting epoch and to immediately start the next voting epoch. Hence, there is no time delay between any two consecutive voting epochs. (3) To thwart the peak-end-effect in voting (see Section~\ref{ssec:Peaknew}),  a variable-time  epoch voting is used. Here, the start time of a new  and upcoming voting epoch is not known in advance and it  cannot be conclusively determined by any of the parties in the election.

	There exist a few issues related to centralized e-voting, such as censorship and tampering with the results and data (see Section~\ref{ssec:blockchainuseAoV}).
	For these reasons, our voting framework may be implemented on a blockchain, such as using Hyperledger projects (see Section~\ref{ssec:permissioned}).
	The bitcoin headers from the Bitcoin network are used as an initial source of randomness (see Section~\ref{ssec:rndnonce}) to trigger the start of the voting epochs. This is unrelated to the blockchain platform on which the voting was implemented.
	Bitcoin headers were chosen because of the difficulty for the mining adversary to find a suitable nonce to the  Bitcoin PoW puzzle within 10 minutes (on average), that  would trigger the voting epoch. 
	
	The VDF is an additional component used in securing the  variable length voting interval from being arbitrarily and intentionally (maliciously)  triggered by the  mining adversary.
	The newly mined blocks from the Bitcoin network are sent to a VDF, to further delay  the mining adversary from  maliciously triggering the voting epoch. 
	Since Bitcoin mining is a lottery, the Bitcoin network accepts the first valid nonce from any miner and appends that mined block to the chain, thereby preventing the malicious miner attack on our voting framework (see Section~\ref{ssec:miningadv} for details).
	The focus of our voting framework is  in thwarting the peak-end-effect and addressing  the security issues due to the introduction of variable time epoch triggers.

\subsection{Design Goals}
\label{ssec:designgoals}

The AoV framework has the following main design goals.
\begin{compactenum}
	\item \textbf{Repeated voting epochs}:  Participants are allowed to continuously vote and change elected candidates or policies without waiting for the next election. Participants are permitted to privately change their vote at any point in time, while the effect of their change is considered rightful at the end of each epoch.
	The duration of such epochs is shorter than the time between the two main elections.
	\item \textbf{Randomized time epochs}: The end of each epoch is randomized and made unpredictable.
	In contrast to fixed-length time epochs, the proposed randomized time epochs are used to thwart the peak-end-effect.
	\item \textbf{Plug \& play voting protocols}: The AoV framework
	is designed to ``plug \& play'' new or existing voting
	protocols. As a result, AoV inherits the properties of
	the underlying protocol chosen.
	However, in the interest of vote confidentiality on a blockchain,  we recommend protocols providing
	secret ballots whose correctness can be publicly verified by $SC$ without leaking any information, e.g., \cite{yu2018platform,ICBC:MRSS19,killerprovotum,Baudron2001}.
	Also, due to the repetitive nature of AoV, e-voting protocols with expensive on-chain computations and required fault recovery (due to stalling participants)  may be less appealing but still acceptable with some limitations, e.g.,~\cite{McCorrySH17,Li2020,DBLP:conf/fc/SeifelnasrGY20,venugopalan2021bbbvoting,stanvcikova2022sbvote} (see also \autoref{ssec:discussion-plugged-protocols}).

\end{compactenum}

\section{Always on Voting Framework}
\label{sec:framework}

Always-on-Voting (AoV) is a framework for blockchain-based e-voting, in which voting does not end when the votes are tallied and the winners are announced.
Instead, participants can continue voting for their previous vote choice or change their vote.
A possible outcome of such repetitive voting is transitioning from a previous winning candidate to a new winner.
To achieve this, the whole time interval between two regularly scheduled elections is unpredictably divided into several intervals, denoted as voting epochs.
Participants may change their vote anytime before the end of a voting epoch (i.e., before a tally of the epoch is computed); however, they do not know beforehand when the end occurs.
Any vote choice that transitioned into the supermajority threshold of votes is declared as the new winner of the election,
and it remains a winning choice until another vote choice reaches a supermajority threshold.

\subsection{Underlying Voting Protocol}\label{sec:underlying-voting-protocol}
AoV provides the option to plug \& play any suitable e-voting protocol.
To provide the baseline security and privacy of votes (with on-chain verifiability), we assume the voting protocol plugged into AoV allows participants to blind or encrypt their votes whose correctness is verified on-chain by $SC$.
However, AoV does not deal with other features supported by the plugged-in voting protocol (such as end-to-end verifiability~\cite{Jonker13}, coercion-resistance~\cite{yu2018platform}, receipt-freeness~\cite{Kiayias2002}, and fairness~\cite{Kiayias2002}).

\subsection{Example of Operation}
\autoref{fig:transition-epoch} illustrates a scenario with 4 candidates $A$-$D$,
where \textit{C} is the present winner of the election.
For example, the supermajority threshold of 70\% votes is set for future winnings, which is a tunable parameter that may be suitably tailored to the situation.
All candidates are initialized to their winning percentages of obtained votes from the last election.
Over time, the individual tally is observed to shift as the supermajority of participants decided to change their vote in favor of another candidate by voting in the epoch intervals.
Through \textit{k} intervals, the winner-ship is seen to transition from candidate \textit{C} to \textit{A}.
At the $\textit{k}^{th}$ interval, \textit{A} obtains the 70\% threshold of votes and is declared as the new winner.
Note that the supermajority is required only in the voting epochs between two regularly scheduled elections.
The regular elections are also executed in AoV, and they repeat every $M$ months/years, while requiring only a majority of votes (i.e., $>$50\%) to declare a winner.
Hence, in contrast to existing electoral systems, we only propose changes between regularly scheduled elections and enable new candidates to be added or removed (see in \autoref{ssec:discussion-removed-part}).

\myparagraph{\textit{Justification for Supermajority}}
A supermajority of 70\% was chosen (see Appendix~\ref{sec:pandr} for background) to help the incumbent carry out reforms without the risk of losing when there is still sufficient support from participants.
On the other hand, the main purpose of this threshold is to block (or repeal) policies that are unpopular or negatively affecting a vast majority of participants.
Further, participants may be inclined to vote in favor of the referendum/proposal if it is coming from a leader who won with a super-majority.
Additionally, we aim to avoid the quorum paradox (see Appendix~\ref{sec:roleAoV})
by setting a minimum participation requirement of 70\% from the just concluded main election.

\begin{figure}
	\centering
	\includegraphics[width=0.80\columnwidth]{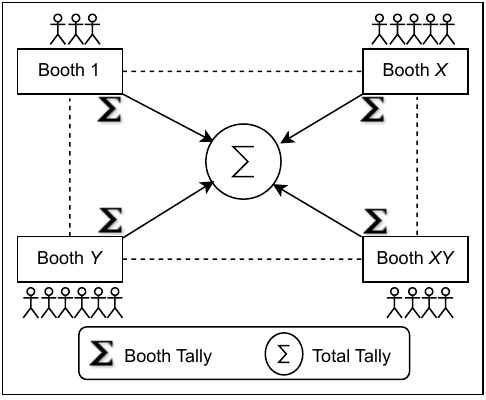}
	\vspace{-0.4cm}
	\caption{When the tally computation is triggered, each booth computes the sum of all votes cast at the booth (referred to as booth tally). Each booth tally is further summed up to determine the total tally.
		Pictorially, the booths are numbered 1 to $X$ along the rows and 1 to $Y$ along the columns. There are a total of $X\cdot Y$ booths.
	}
	\label{fig:booth}
	\vspace{-0.4cm}
\end{figure}

\subsection{Overview of AoV Phases}
\label{ssec:votingphases}
Once the setup phase (that ensures participants agree upon all system parameters) is completed, electronic voting frameworks typically consist of three phases:
(1) a registration phase to verify voter credentials and add them to the voting system,  
(2) a voting phase, in which participants cast their vote via a secret ballot, and
(3) a tally phase, where the total votes for each candidate are counted and revealed to participants.
The voting protocol plugged-in with the AoV framework may contain additional phases, but we omit them here for brevity.

\begin{figure}
	\centering
	\includegraphics[width=0.9\columnwidth]{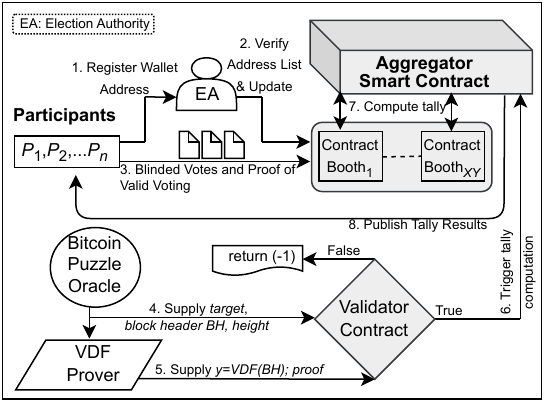}
	\vspace{-0.3cm}
	\caption{Interaction among participants ($P$s), election authority ($EA$), smart contracts, the Oracle, and VDF prover.
		(1) Registering wallet addresses of participants and (2) their identity verification are made by the $EA$.
		(3) Participants send a blinded vote and   proof of valid voting to their assigned booth contract.     
		The booth contract verifies the validity of the vote.
		(4) The Bitcoin Puzzle Oracle (BPO) provides the latest Bitcoin block header (BH) and
		(5) VDF prover sends a proof of sequential work with $y$ (the output of VDF(BH)) to the validator contract.
		(6) The validator contract finishes the epoch and shifts the state of the elections to the tally upon meeting the required conditions.
		(7) The aggregator contract is responsible for totaling individual booth tallies and (8) publicly announcing the total tally.  The on-chain components of AoV are depicted in gray.}
	\label{fig:model}
	\vspace{-0.4cm}
\end{figure}

The architecture of AoV is shown in \autoref{fig:model}.
In AoV, participants (in step 1) register their wallet address\footnote{Refer to Appendix~\ref{ssec:appendixWallet} for a proposed method to improve anonymity for wallet users.} with the EA, who then (in step 2) verifies and updates it on the booth smart contract\footnote{Participants are randomly grouped and assigned to booths  $\in\{1,2,...,X\cdot Y\}$ (see \autoref{fig:booth}), represented by a booth smart contract.}.
This is followed by the voting phase (in step 3), where participants cast their secret vote (i.e, not revealing the vote choice).
The BPO (step 4)  supplies the validator contract and VDF prover with the $target$, recent
Bitcoin block header $BH$ and its block height.
The VDF prover\footnote{A VDF prover is any benign user in the voting ecosystem with commercial hardware to evaluate the input of VDF, i.e., $y~=~$VDF($BH$) and supply a proof $\pi$.
} (in step 5) computes and submits   $VDF(BH)$ and a  proof of sequential work ($\pi$) to the validator contract.
The validator contract (in step 6) verifies the VDF proof and checks whether the supplied nonce $s$ (included in the block header) is a valid solution to the Bitcoin PoW puzzle of the supplied header.
If both verifications pass and a certain equation outputs to 0 (more details in Section~\ref{ssec:interactions}),
the validator contract finalizes the epoch and triggers the tally computation for the epoch.
Otherwise, it waits for the next block header submission from the BPO and the proof of sequential work from the VDF prover.
When the tally computation is triggered, each booth contract $\{1,2,...,X\cdot Y\}$, sums up all its local vote counts and sends them to the aggregator contract (step 7).
Then, the aggregator contract totals the votes from each booth contract\footnote{Refer to Appendix~\ref{ssec:appendixSharding} for privacy implications of booth sharding.} and publishes the final tally (step 8).

In AoV, the $EA$ is authorized to register/remove participants and candidates in a future interval.
Nevertheless, candidates can also be managed by other means, and AoV does not mandate how it should be done (see discussion in \autoref{ssec:discussion-removed-part}).
When there are no other changes in the next interval, revoting repeats with step~2 and ends with step~8.

From the initialization of AoV until the next regular elections, the validator smart contract accepts all future Bitcoin block headers.
The new block headers (as part of their blocks) arriving every 10 minutes on average are appended to the Bitcoin blockchain.
The BPO is responsible for timely supplying\footnote{
	To respect the finality of the Bitcoin network, we assume that BPO supplies only the block headers that contain at least 6 confirmations on top of them. As a consequence, the     probability that such a confirmed block will be reverted is negligible. Note that this does not influence the chances of $ADV_{min}$ to succeed since she is already  ``delayed'' by VDF in finding multiple PoW solutions at the same height; therefore, she prefers to work on top of the chain with her new attempts. }    
each new block header to the VDF prover and validator contract.
The VDF prover computes the VDF on each of those block headers after they are supplied.

\subsection{Calculating the Epoch Tally Time}
\label{ssec:vrf}
Due to concerns that Bitcoin nonces are a weak entropy source, additional steps are taken to make it cryptographically secure (see details in  \autoref{ssec:rndnonce}).
Our notion of randomness relies on Bitcoin Proof-of-Work to generate valid nonces.\footnote{If the nonce overflows, a parameter called \textit{extraNonce}  (part of the coinbase transaction) is used to provide miners with the extra entropy needed to solve the PoW puzzle.}
The validator contract awaits future block headers yet to be mined on the Bitcoin network.
When new Bitcoin block headers arrive, they are sent to the validator contract and the VDF prover via the BPO.
The VDF ensures that a mining adversary cannot find more than one valid nonce to the block at a given height and test if the nonce is favorable within 10 minutes.
The VDF is computed with the block header at the input by the VDF prover, who then submits the VDF output and the proof of sequential work to the validator contract.
The choice of VDF depends on its security properties, speed of verification, and a size of the proof~\cite{Pietrzak2019}.  
Let  $BH$ be the Bitcoin block header.
Once VDF prover computes $y=VDF(BH)$, a small proof ($\pi$) is used to trivially verify its correctness using $VDF\_Verify(y,\pi)$. Wesolowski’s construction~\cite{Wesolowski2020} is known for its fast verification and a short proof:
Let $TL$ be the number of sequential computations.
Prover claims
$$y~=~BH^{2^{TL}}$$
and computes a proof $$\pi= BH^{\floor{\frac{2^{TL}}{B}}},$$ where $B$~=~Blake256($BH~||~y~||~TL$) hash.  
Verifier checks whether $$\pi^{B}\cdot BH^{2^{TL}mod\; B}\stackrel{?}{=} y.$$
Since we employ VDF, $Adv_{min}$ does not know the value of $y$ before evaluating the VDF and is forced to wait for a given amount of time to see if the output is in her favor (before trying again).
However, since Bitcoin mining is a lottery, other miners can solve the puzzle and append a block by propagating the solution to the Bitcoin network, rendering any withheld or attempted solution by the adversary that was not published useless.

\medskip
\subsubsection{Interactions of BPO, VDF Prover, and Validator}
\label{ssec:interactions}
Let $TotalTime$ be the time in minutes between 2 regular elections.
The BPO (see step 4 in \autoref{fig:model}) feeds the block header $BH$ of every future Bitcoin block (when it is available) to the validator contract and VDF prover.
Further, BPO provides the validator contract  with the value of $target$ when it changes; i.e., every 2016 block (see \autoref{ssec:puzzlepow}).

Upon obtaining data from BPO, the VDF prover computes VDF output  \begin{equation}\label{eqn_vdfprover}
	y~=~VDF(BH)
\end{equation}
with the VDF proof $\pi$ and sends them to the validator contract (see step 5 in \autoref{fig:model}).
Next, the validator contract verifies the following conditions:
\begin{equation}\label{eqn_vdfverifier}
	VDF\_Verify (y,\pi) \stackrel{?}{=} True,
\end{equation}
\begin{equation}\label{eqn_0}
	SHA256(BH)<target.
\end{equation}
The first verification checks whether the VDF output $y$ and supplied proof $\pi$ (i.e., \autoref{eqn_vdfverifier}) correspond to the BPO-supplied block header $BH$.
The second verification (i.e., \autoref{eqn_0}) checks whether the nonce received from BPO\footnote{We note that the BPO may be replaced by a quorum to improve decentralization. The validator contract will then accept the input from BPO only when $2/3$ (and more) of the quorum is in agreement.}
is a valid solution to the Bitcoin PoW puzzle.
Once both checks pass, the validator contract proceeds to compute
\begin{equation}\label{eqn_1}
	a=SHA(y),
\end{equation}
where SHA(.) is SHA-X-256\footnote{X denotes a suitable hash function such as SHA-3, and 256 is the output length in bits.}  hash.
The goal of \autoref{eqn_1} is to consolidate the entropy by passing it through a compression function that acts as a randomness extractor (see \autoref{ssec:rndnonce}).
Using $a$, the validator contract computes

\begin{equation}\label{eqn_2}
	b= a\; (mod\; BHsInInterval),
\end{equation}

where the expected number of block headers are
\begin{equation}\label{BHInInterval}
	BHsInInterval ~=~ \frac{IntervalTime}{BlockTime}.
\end{equation}

and the time interval is found as
\begin{equation}\label{eqn_3}
	IntervalTime ~=~ \frac{TotalTime}{ft}.
\end{equation}

As seen in \autoref{fig:transition-epoch}, $ft$ is the number of intervals (epochs) that the total time ($TotalTime$) between 2 regular elections is divided into. $BlockTime$ is the average time of block generation (i.e., 10 minutes in Bitcoin).

The end  for the current voting interval and  computation of its tally is triggered when the output of the validator contract in Equation~\ref{eqn_4} is $True$ (see step 6 in \autoref{fig:model}):
\begin{equation}\label{eqn_4}
	VC_{output} = \begin{cases} \mbox{True,} & \mbox{if } b = 0 \\ \mbox{False,} & \mbox{otherwise}. \end{cases}
\end{equation}

\noindent
\paragraph*{\textbf{Example}}
Let $TotalTime$ = 4 years = $525600\cdot 4$ minutes and $ft=8$; then $IntervalTime = (525600\cdot 4)/(8)= 262800$ minutes $\approx 182.5$ days and $BHsInInterval$ = $262800 / 10$ = 26280 blocks.
Therefore, the BPO will send on average 26280 block headers ($BH$ values) to the validator contract within an assumed 182.5 days long epoch (assuming 10 minutes block creation interval), i.e., $1/8$ of the total time.
We expect the tally will be triggered on average once in every 182.5 days because of the Poisson probability distribution of this event.
Therefore, $ft$ expresses the expected number of epochs, while $ft$ might differ across the regular election iterations.



\section{Analysis}
\label{sec:analysis}

\subsection{Mining Adversary}
\label{ssec:miningadv}

The goal of mining adversary $Adv_{min}$ is to find a valid nonce $s$ that solves the Bitcoin puzzle such that $b$ in \autoref{eqn_2} is 0. When these two conditions are met, it marks the end of the current voting epoch and the validator contract triggers the tally computation of votes for the just finished epoch.
We set the difficulty for the benign VDF prover (with commercial hardware) to take 100 minutes\footnote{ We consider $A_{max}=10$, i.e., what is solved by a benign VDF prover in 10 units of time, while in the case of $Adv_{min}$ it is in 1 unit.} to solve VDF($BH$). Based on $A_{max}$ limit (see Section~\ref{ssec:VDF}), we assume $Adv_{min}$ to take at least 10 minutes to solve the VDF. As a result, $Adv_{min}$ is restricted to a maximum of 1 try (considering 10 minutes as an average Bitcoin block creation time), excluding the Proof-of-Work required to solve the Bitcoin mining puzzle.  During this time, another  miner would have solved the Bitcoin PoW puzzle, propagated the corresponding block to the Bitcoin network, and appended it to the chain.
	The additional workload from the VDF denies a mining adversary the time required to find a suitable nonce that triggers the end of current voting epoch and send that corresponding block to the Bitcoin network for inclusion.
	In Figure~\ref{fig:model}, we reiterate that the Bitcoin Puzzle Oracle only supplies the block headers fetched from the Bitcoin network.

However, since a Bitcoin block header is generated on average once every 10 minutes and the benign VDF prover is occupied for 100 minutes, the question is -- how many VDF provers are required to prevent the block headers from queuing up?
We can see in \autoref{tab:vdftimeline} that VDF prover 1 runs a task for time 0-100 minutes, and she picks up the next task to run for time 100-199 minutes. Similarly, all other provers pick up the next task after completing the present one.
Hence, 10 VDF provers are sufficient to prevent block headers from queuing up because $A_{max}=10$.

On the other hand, a benign VDF prover might reduce $A_{max}$ of VDF computation by using specialized hardware instead of commercial hardware (depending on the cost-to-benefit ratio).  
However, we emphasize that the VDF can be computed only after solving the PoW mining puzzle, which is prohibitively expensive. Moreover, the puzzle difficulty increases proportionally to the mining power of the Bitcoin network.
Hence, the proposed serial combination of solving the Bitcoin mining puzzle followed by the computation of VDF output improves the aggregate security against $Adv_{min}$ from choosing a favorable nonce.
The estimated requirement of $A_{max}=10$ might be further increased as more studies to efficiently solve VDFs on ASICs are carried out.
However, if $A_{max}$ will increase in the future, our solution can cope with it by employing more VDF provers.

\begin{table}[t]
	\centering    
	\begin{tabular}{ p{1cm} p{1.9cm} | p{1cm} p{1.9cm}  }

		\toprule
		VDF Prover& Time (minutes) & VDF Prover&Time (minutes)\\
		\midrule
		1   & 0-100	&  1   & 100-199\\
		2   & 10-110  &  2   & 110-209\\
		3   & 20-120  &  3   & 120-219\\
		4   & 30-130  &  4   & 130-229\\
		5   & 40-140  &  5   & 140-239\\
		6   & 50-150  &  6   & 150-249\\
		7   & 60-160  &  7   & 160-259\\
		8   & 70-170  &  8   & 170-269\\
		9   & 80-180  &  9   & 180-279\\
		10  & 90-190  & 10   & 190-289\\    
		\bottomrule
	\end{tabular}
	\vspace{0.2cm}
	\caption{\label{tab:vdftimeline}Scheduling 10 VDF provers without queuing. Note the VDF computations on a VDF prover machine are not parallelized. It is the scheduling alone that is in parallel. The start time is based on the job arrival time at the VDF prover, where it will run for 100 minutes. Once completed, it is ready to take on the next job. In column 2, the start times are 10 minutes apart and correspond to the average BTC interblock (job) arrival time. The largest idle time in column 1 is for VDF Prover 10 at 90 minutes, waiting for the job to start. Beyond this, all VDF prover machines are continuously occupied since a new job is available to start immediately after the current job ends. }    
	\vspace{-0.6cm}
\end{table}

\subsection{Implications of VDF Prover Synchronization and Optimizing Frequency of Supplied Block Headers}
\label{ssec:bh}
Several VDF provers are  synchronized to supply the VDF proofs to the validator contract in sequence.
However, there are no adverse effects when the proofs are generated and supplied out of sequence.
The validator smart contract stores the latest \textit{block height} for which the VDF proof was last  accepted. It only allows  proof verification for stored \textit{block height}$+1$ on the contract and any out-of-order proofs have to be re-sent.
Once the  order  is corrected, a handful of VDF proofs may appear in quick succession at the validator contract.
However, the tally for the interval is only triggered when $VC_{output}$ in \autoref{eqn_4} is \textit{True}.

In terms of gas consumption, it can be costly to process every single Bitcoin block header (supplied to the VDF prover and the validator contract by the BPO).
We suggest optimizing this by choosing a coarser time granularity of the block header supply, independent of the Bitcoin block interval (e.g., every x-th block).
We modify the example from \autoref{ssec:interactions} by considering the processing of every 100$^{th}$ Bitcoin block header.\footnote{Note that this would need another condition to be met, i.e., the block height (mod 100) should be equal to 0.}
$TotalTime$ = 4 years = $525600\cdot 4$ minutes and the total number of intervals $ft=8$.
Then, $IntervalTime = (525600\cdot 4)/(8) = 262800$ minutes and $BHsInInterval = 262800/(10\cdot 100) = 262.8$.
On average, the oracle will send 262.8 block headers ($BH$ values) to the validator contract within 182.5 days instead of the 26280 block headers required in the original example.
In this case, we only need 1 VDF prover instead of 10, and it provides similar security guarantees as before.

\subsection{Randomness of Bitcoin Nonces \& AoV Entropy}
\label{ssec:rndnonce}
We decided to utilize a single public source of randomness due to the low computation cost incurred to us, instead of a distributed randomness  with higher synchronization complexity.
Bonneau et al.~\cite{BonneauCG15} showed that if the underlying hash function used to solve the Bitcoin PoW puzzle is secure, then each block in the canonical chain
has a computational min-entropy of at least $d$ bits, representing the mining difficulty.
I.e., $d$ consecutive 0 bits must appear in the hash of the block header\footnote{At the time of writing, $d\approx$ 76.}.
Hence, $\floor{\frac{d}{2}}$ near-uniform bits can be securely extracted.
Nevertheless, empirical evaluation has shown that Bitcoin nonces have visible white spaces (non-uniformity) in its scatter-plot~\cite{Bitmex2019}.
A possible explanation is that some miners are presetting some of the bits in the 32-bit $nonce$ field and using the $extraNonce$ to solve the PoW puzzle.
We use the entire block header as the initial source of entropy instead of the 32-bit nonce alone to avoid such biases.
To reduce the probability of $Adv_{min}$ biasing the solution in her favor, the block header is passed through a verifiable delay function (see \autoref{eqn_vdfprover}).
The output of VDF is hashed (see \autoref{eqn_1}) to consolidate the entropy.

\section{Discussion}
\label{sec:discussion}

\subsection{Voting in a Referendum}
\label{ssec:referendum}
The proposed framework for repetitive voting is also suitable for a referendum, where participants may vote on a proposal.
Unlike elections scheduled at regular intervals of time, a referendum may not be necessarily tabled (i.e., put up for voting) more than once.
Nevertheless, AoV may be used to make the referendum voting repetitive.
Any change to the outcome of a referendum using the AoV framework requires a supermajority unless the proposal is re-tabled through an agreement, in which case normal operations follow.\footnote{Some referendums require only a majority while others need a supermajority to arrive at a decision. AoV does not change the voting requirements when it is first tabled or re-tabled.}
The danger of low participant turnout in-between regularly scheduled votings may be mitigated by setting a minimum threshold on the number of participants (e.g., 70\% of the just concluded elections) to overturn a decision.
In the case of a referendum, a similar minimum threshold based on previous participant turnouts is recommended.
Further, the proposed framework may also be used as an extra tool to record changing public opinion.

\subsection{Voting Costs on a Public Permissioned Blockchain \& Incentives for the VDF Prover}
\label{ssec:permissioned}
The expenses imposed by a public permissionless smart contract platform may be high. 
Furthermore, the transactional throughput of such platforms may be insufficient to cater to a larger number of participants voting in a specified time window.
To reduce costs and improve performance, AoV can run on a public permissioned Proof-of-Authority (PoA)\footnote{In a PoA consensus,  a strict vetting process is used for machines that are given the right to generate blocks. Vetting is carried out by pre-approved moderators who checks the blocks and its transactions.} blockchain, e.g., using Hyperledger projects (such as safety-favored Besu with BFT), in which all nodes have the same consensus power.
Since nodes in PoA protocols ``stake'' their reputation that is backed by the knowledge of their identities, any misbehavior can lead to loss of the reputation, which is expensive and nodes are thus naturally disincentivized from such misbehaviors.
Optionally,  smart contract platforms backed by trusted computing that off-chains expensive computations may be used (e.g., Ekiden~\cite{cheng2019ekiden} and TeeChain~\cite{lind2019teechain}). Other partially-decentralized second layer solutions (e.g., Plasma\footnote{See \url{https://plasma.io/plasma.pdf}.}, Polygon Matic,\footnote{See  \url{https://github.com/maticnetwork/whitepaper}.} and Hydra\footnote{See  \url{https://hydra.family/head-protocol/}.}  may also be used. 
Even though these solutions might preserve most of the blockchain features harnessed in e-voting, availability and decentralization may be decreased. 
The selection depends on the security/performance trade-off (alike the number of full nodes used).

In a permissioned blockchain, only authorized participants are allowed to append transactions to the chain.
Unlike a public blockchain, the only costs involved in a permissioned blockchain reside in maintaining the blockchain infrastructure.
Consider the following example to put the costs of the PoA blockchain into the real perspective. 
Let us assume that 4 independent candidates are contesting the election, along with 6 public notaries and 1 $EA$.
Each of them may run a consensus node or rent it out from a cloud for a low fee (e.g., $\sim$20-50 USD/month). The consensus nodes might be run by $EA$, public notaries, and some of the participants.
The VDF provers may be incentivized through a cryptocurrency\footnote{Funds may also be supplied via central bank digital currency.} treasury~\cite{ZhangOB19} to supply crypto-coins for their services (supplying VDF proofs). A fraction of all transaction fees on the blockchain may be sent to a treasury smart contract and claimed by the VDF prover upon verification of a valid VDF proof by the validator contract in \autoref{fig:model}.
This incentive would ensure a consistent supply of the VDF proofs, which is required to maintain the probability of entering a new epoch within each of the predefined intervals.

\subsection{Limitations}
\label{ssec:limitations}
AoV may be less suitable for voting where a winner is found through set reduction\footnote{In set reduction, candidates are eliminated based on a tally cut-off criteria and revote is carried out on the reduced candidate set.}.
AoV is suited for elections with 1-out-of-$k$  voting choices, i.e., plurality voting such as first-past-the-post (FPTP)~\cite{Harewood2003}.
Also, FPTP is the second most commonly used electoral system in the world. It covers 59/232 countries in national legislature elections and 24/232 countries in presidential elections~\cite{Idea2021}, encompassing billions of voting participants under this electoral system.

\subsection{Blockchain in the AoV Framework}
\label{ssec:blockchainuseAoV}
In our approach, the blockchain with smart contract platform enables us to enforce the rules of the elections, including the plugged-in e-voting protocol as well as wrapping the AoV framework itself.
In the case of plugged-in protocol, the blockchain verifies the correctness of submitted (private) votes, computes the tally in a verifiable fashion (i.e., tallied-as-recorded verifiability), and enables participants to achieve cast-as-intended and recorded-as-cast verifiability, all together providing end-to-end verifiability~\cite{benaloh2015end, killerprovotum}.
Furthermore, in the case of AoV that wraps the plugged-in blockchain-based e-voting protocols, the blockchain enables  to re-vote anytime and the correctness of their vote is verified by a smart contract.

\subsection{Randomization of the Epoch End Times}
The voting intervals in AoV have soft bounds on when the end of an epoch is triggered, i.e., it triggers when $a$ in \autoref{eqn_1} has a zero remainder when divided by $BHsInInterval$ in \autoref{eqn_2}. 
In particular, the event of interval end follows the Poisson distribution.
The end of the interval is triggered faster when the value of $IntervalTime$ is smaller.

To the extent, indecisive voters may be prompted and reminded to vote (e.g., by push notifications in a smartphone, or emails). 
For example, a smartphone or tablet may prompt the user with a push notification at the start of each new epoch. 
It may also send reminders that they had not already voted in that epoch after a certain elapsed time. 
If a voter does not have a smartphone, there may be alternative solutions for notification, such as sending an email.

\subsection{Adding and Removing Candidates}
\label{ssec:discussion-removed-part}
AoV allows the list of candidates to be updated at any time by the authorized entity (or entities), and thus candidates not willing to participate can be removed in the new epoch. Similarly, new candidates can be added.
Such rules may be  added to the smart contract based on the agreement between the $EA$ and the candidate (or even in some decentralized fashion).
In the context of this work, we abstract from the implementation details of this aspect, and we focus on addressing the continuous voting and peak-end effect.

\section{Related Work}
\label{sec:related}

\subsection{Blockchain-Based E-Voting}
\label{ssec:blockchainevotingrelatedwork}

Many blockchain-based e-voting protocols and systems are present in the literature, 
out of which, we are not aware of any continuous e-voting system that addresses the peak-end effect problem.
In the following, we extend the categorization of remote blockchain-based e-voting systems proposed by Yu et al.~\cite{yu2018platform} with more examples.

\subsubsection{Voting Systems Using Smart Contracts}
\label{ssec:related-sc}
McCorry et al.~\cite{McCorrySH17} proposed OVN, a self-tallying voting protocol that provides vote privacy and supports two vote choices; however, it does not provide robustness (i.e., recovery from stalling participants) and suffers from expensive  computation on the smart contract.
A similar approach based on the same protocol was proposed by Li et al.~\cite{Li2020}, who further provided robustness (based on Khander et al.~\cite{KhaderSRH12}).
Seifelnasr et al.~\cite{DBLP:conf/fc/SeifelnasrGY20} increased the scalability of OVN by off-chaining tally computation and registration at $EA$. 
Due to the higher costs imposed by storing data on smart contracts, they compute the Merkle tree of voter identities and store only its root hash in the smart contract. 
Their approach requires only a single honest participant to maintain the protocol's security.
Venugopalan et al.~\cite{venugopalan2021bbbvoting} proposed BBB-Voting, an approach that on top of OVN features enables $k \geq 2$ vote choices and provides robustness and further cost optimizations.
Stan{\v{c}}{\'\i}kov{\'a} and Homoliak~\cite{stanvcikova2022sbvote} proposed SBvote, an approach that  extends BBB-Voting by integrity-preserving batching and hierarchical booth aggregation to improve scalability, which supports millions of participants.  
Yu et al.~\cite{yu2018platform} employed ring signatures to ensure that the vote is from one of the valid choices, and they achieve scalability by linkable ring signature key accumulation. 
Their approach provided receipt-freeness 
under the assumption of trusted $EA$;
however, it does not provide end-to-end (E2E) verifiability. 
Killer et al.~\cite{killerprovotum} presented Provotum, an E2E verifiable remote voting system with 2 vote choices. 
The authors employed threshold cryptography to achieve robustness using a scheme similar to Shamir secret sharing. 
Matile et al.~\cite{ICBC:MRSS19} proposed a voting system providing cast-as-intended (but neither  E2E nor universal) verifiability. 
Their system uses ElGamal encryption based on DLP with integers modulo $p$. 

\subsubsection{Voting Systems Using Cryptocurrency}
Zhao and Chan~\cite{zhao2015vote} proposed a privacy-preserving voting system with 2 vote choices based on Bitcoin, which uses a lottery-based approach with an off-chain distribution of voters' secret random numbers. 
Random numbers enable to hide the vote choice and are distributed via ZKP.
The authors use deposits to  incentivize participants to comply with the protocol. 
Tarasov and Tewari~\cite{tarasov2017internet} proposed a conceptual voting system based on Zcash. 
The voter's anonymity (and thus the privacy of vote) is ensured by the z-address that preserves unlinkability. 
The correctness of the voting is guaranteed by the trusted $EA$ and the candidates. 
Liu and Wang~\cite{liu2017voting} proposed a conceptual voting approach that is based on blind signatures with 2 vote choices. 
They utilized blockchain only for auditable sending of messages among participants and $EA$.

\subsubsection{Commercial Voting Systems with Ballot Box}
This category of voting systems usually does not offer the privacy of votes, instead relies on the unlinkability of blockchain wallet addresses with participant identities; therefore, one may use VPNs and/or anonymization services.
Examples are FollowMyVote\footnote{See \url{https://followmyvote.com/}.}, Tivi,\footnote{See \url{https://tivi.io/tivi/}.},  and Agora,\footnote{See \url{https://www.agora.vote/}.} 
which use blockchain only for recording votes into a \textit{ballot box}.  
In contrast to the previous solutions, NetVote\footnote{See \url{https://github.com/netvote/elections-solidity}.} is a solution addressing privacy by trusting $EA$ to reveal the encryption key of ballots after the elections end.

	\subsection{Plug-able Protocols into AoV}
	\label{ssec:discussion-plugged-protocols}	
	We reviewed several blockchain-based e-voting approaches.  
	The most  suitable approaches to be plugged in AoV are the ones that allow re-voting within an epoch (i.e., participants can change their votes), such as~\cite{yu2018platform,ICBC:MRSS19,killerprovotum,Baudron2001}.
	However, approaches using distributed MPC keys, a.k.a., self-tallying protocols~\cite{Kiayias2002} such as~\cite{McCorrySH17,Li2020,DBLP:conf/fc/SeifelnasrGY20,venugopalan2021bbbvoting,stanvcikova2022sbvote} have a limitation in terms of AoV despite having other benefits (e.g., perfect ballot secrecy, fairness).
	In particular, they can be applied in AoV with a limitation of a single vote within an epoch.
	They do not enable re-voting within an epoch since they provide the privacy of vote only with a single cast of a vote (i.e., requirements on the shared MPC key). With shared MPC keys, the process of removing stalling voters requires additional steps, before the tally of votes can be computed. The voting protocol plugged-in to AoV is left to the implementer based on their exact needs.

\section{Conclusions}
\label{sec:conclusions}
We reviewed the works in political science that motivated an engineering solution to the peak-end effect problem in voting and voter manipulation.
We showed that existing voting systems provide little to no recourse with changing the elected candidates (until the next main elections) -- even when they lost the support of a majority of voters.
Therefore, we proposed the Always-on-Voting (AoV) framework that allows participants to change their vote between two main elections and thwart the peak-end effects.
To achieve this in unbiased fashion, we divided the time between two main elections into a few shorter epochs whose ends were made unpredictable, and tallying the votes at the end of each epoch.
The AoV framework used  
verifiable delay functions and Bitcoin block headers as the source of randomness to thwart the mining adversary who intends to bias the ends of epochs.
AoV recommended a public permissioned blockchain to ensure its security properties and save costs.
It can be integrated with various existing blockchain-based e-voting solutions.
Finally, we proposed a supermajority requirement to elicit a change reflecting the current result of the elections, which helps to maintain the stability of the existing system.

\section*{Acknowledgment}
The authors would like to thank Pawe\l{} Sza\l{}achowski, Stefanos Leonardos, Daniel Reijsbergen, Mike Rosing, and Balamurali B. T. for their comments on an earlier version of the paper. 
Ivana Stan\v{c}\'{i}kov\'{a} and Ivan Homoliak were supported by the DIAS project, which has received funding from the ECSEL Joint Undertaking under grant agreement No 8A21012, 101007273.
Any opinions, findings and conclusions or recommendations expressed in this material are those of the author(s) and do not reflect the views of their employer(s).

\bibliographystyle{IEEEtran}
\bibliography{ref-orig1}
\vspace{0.5cm}The authors are Sarad Venugopalan,
Ivana Stan\v{c}\'{i}kov\'{a} and Ivan Homoliak -- biography not provided by the author's choice.



%






\clearpage
\appendices


\section{Spatial Model \& Median Voters Theory}
\label{sec:pandr}
\begin{figure}[t]
	\centering
	\includegraphics[width=1.0\linewidth]{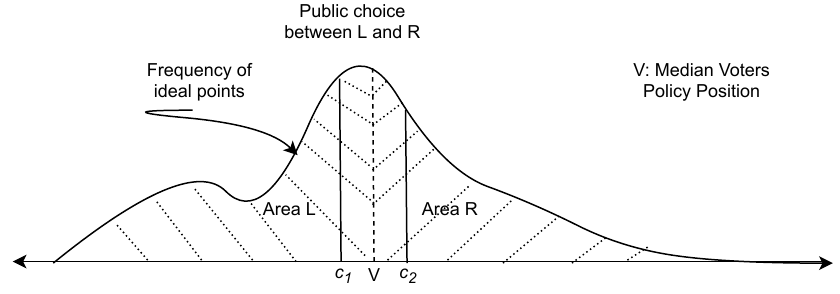}
	\caption{Public choices in voting.
		When candidate $c_1$'s ideal appearing to the left of $V$ and the candidate $c_2$'s  ideal  to the right of $V$ are equidistant from the median voter policy position; $Area\; L\; =\; Area\; R.$}
	\label{fig:publicchoices-median-1}
\end{figure}

First, we discuss the stakeholders, the model used, and the equilibrium of the policies w.r.t. voters. Further, we see the effects when candidate policies shift away from the median voters' policy.
Consider two sets of parties, one the incumbent and the opposition.
Both incumbent and the opposition parties may or may not consist of one or more alliance parties with moderately similar ideologies/goals.
Let $c_1$ be a candidate from the incumbent party and $c_2$ be a candidate from the opposition (see \autoref{fig:publicchoices-median-1}).
We use the spatial voting model, where policy alternatives can be represented in a geometric space (e.g., a 2-dimensional plane), where voters have preferences defined over the alternatives. The spatial model assumes single-peakedness, i.e., the voters' most preferred policy outcome can be represented by a single point. This is called the ideal point of the voter. It implies that overall policy preferences can be marked in increasing order. The frequency of an ideal is the number of voters who scored the same ideal.
The median voter policy position (point V) is where the median line (vertical line at V) divides the area under the graph into 2 equal halves (where Area L = Area R), each with the same number of voters (see \autoref{fig:publicchoices-median-1}).
Further, the spatial model is also considered to be symmetric. I.e., if a voter's ideal point is in between the ideal of candidates $c_1$ and $c_2$, that voter has no preference for one over the other. Since both are equidistant from the voter ideal (which in the illustration shown in \autoref{fig:publicchoices-median-1} is also at the median voter's ideal), half the votes will go to $c_1$ and the other half to $c_2$.

\begin{figure}[t]
	\centering
	\includegraphics[width=1.0\linewidth]{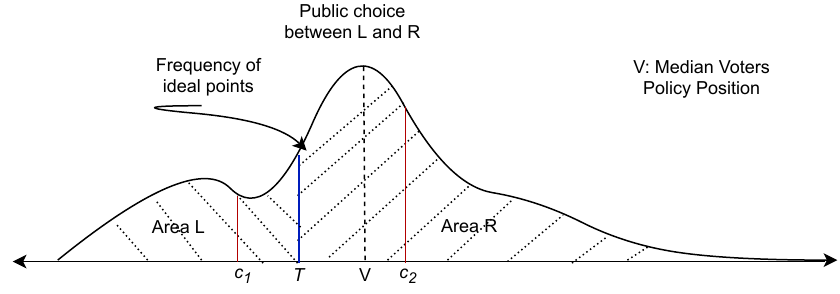}
	\caption{Public choices in voting.
		When the ideal of candidate $c_1$ moves towards the far left, it is farther away from the median voters' ideal. Since voters end up voting for the candidate closer to their ideal point, the $Area\; R\; >>\; Area\; L$, and may result in $c_1$ losing the election.}
	\label{fig:publicchoices-median-2}
\end{figure}

We use the median voters theory\footnote{See \url{https://en.wikipedia.org/wiki/Median\_voter\_theorem}.} as a means to determine the demand aggregation of voters' preferences based on their ideal. According to it, the candidate closest to the viewpoint of the median voter will win. The median voter ideal point is also in equilibrium and stable.
In \autoref{fig:publicchoices-median-1}, since both $c_1$ and $c_2$ are equidistant from the median, there is no clear winner.
However in \autoref{fig:publicchoices-median-2}, candidate $c_1$ has an ideal to the left of the median. Let $T$ be the ideal point equidistant from $c_1$ and $c_2$. Then, half the candidates in the area enclosed between $c_1$ and $T$ will vote for $c_1$, including everyone with ideals to the left of $c_1$'s ideal. The other half of voters between $T$ and $c_2$, including everyone to the right of $c_2$'s ideal, will vote for $c_2$. As a result, $c_2$ will win the majority of votes.
Note the area under the curves in \autoref{fig:publicchoices-median-1}, and \autoref{fig:publicchoices-median-2} may be of any shape and are plotted to illustrate the spatial model in two dimensions.

\section{Role of AoV}
\label{sec:roleAoV}

The role of AoV is to push elected candidates whose ideals (over time) move away from the voters' median ideal back towards equilibrium if they want to remain the winner. This is achieved by allowing voters to vote repeatedly in between 2 main elections.
A supermajority threshold to change an elected candidate is provided to ensure stability when the incumbent candidate pursues major reforms.
However, when radical reforms from the incumbent are backed only by a minority of the voters, the candidates' ideal moves farther away from the median. 
Further, there are costs incurred to a voter when the candidate's ideal moves away from her ideal. When a candidate's ideal is far from the median, the costs to a majority of voters (and hence to society) are high.
Using AoV, when the supermajority winning threshold is met (e.g., 70\%) --- it allows the incumbent to be replaced with a new winning candidate. 
Further, AoV requires a high participant turnout (e.g., 70\%  voters from the concluded main election) for the votes to be tallied in the interval. In some cases, it might result in the quorum paradox, where participants with allegiance to a party are asked to abstain from voting, so the minimum participant threshold is not met.
The quorum paradox is less likely to happen with AoV, especially with a high winning threshold.
For example, in AoV, we set minimum participants = 70\% and winning threshold = 70\%.
If 30\% of participants (from the eligible population) abstain, in theory, the other party may get 100\% of the votes and win the election.
However, if either more than $30\%$ of the voter abstain or more than $30\%$ vote a ``\textit{No}", they are of the same consequence. It is no longer possible to breach the winning supermajority threshold.

\section{Anonymizing Identity}
\label{ssec:appendixWallet}

\begin{figure*}[t]
	\centering
	
	\subfloat[][Booth  with 30 participants, candidate winning \ensuremath{p=0.9}.]{
		\includegraphics[width=0.5\textwidth]{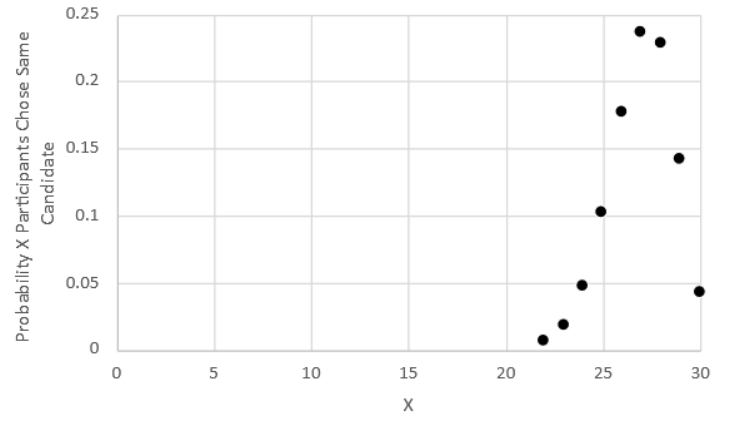}
	}
	\subfloat[][Booth  with 100 participants, candidate winning \ensuremath{p=0.9}.]{   	 
		\includegraphics[width=0.5\textwidth]{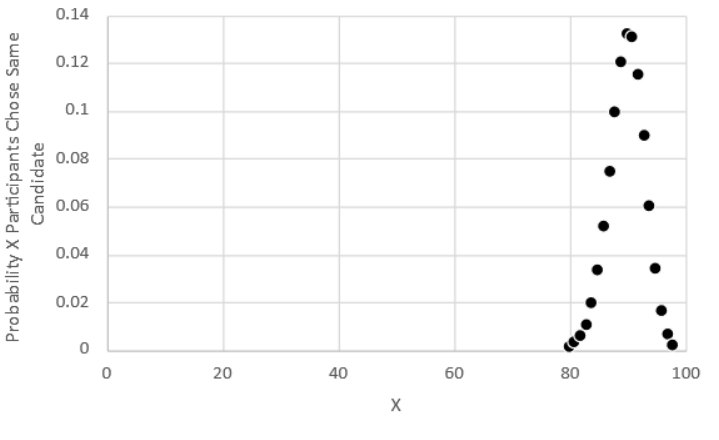}   	 
	}

	\caption{Binomial probability distribution function of X booth participants voting for their favorite candidate whose winning probability is $p$.}
	
	\label{fig:binomialpdf}
	\vspace{-0.4cm}
\end{figure*}

In our voting scheme, we  employed wallet addresses (see Fig.~\ref{fig:model}) to keep track of authorized participants.
A wallet may be considered as a unique pseudo-random string that  points to its owner (voting participant).    
However, this might result in an adversary making statistical inferences by attempting to map the user voting pattern to her wallet address.    
For this specific case, we introduce a new network adversary $Adv_{net}$ who is a passive listener of communication entering the blockchain network but cannot block it.
Her objective is to derive statistical inferences determining the voting patterns of participants (including the voting intervals in which they voted).

The map between participant $P$ and her wallet, recorded by the $EA$ is to prevent sybil attack (preventing any unauthorized person from voting) and double voting. For this reason, $EA$ is trusted to keep this mapping private. Only voters corresponding to white-listed wallets by $EA$ are allowed to vote and everyone else is blocked from voting on the smart contract. The wallet address corresponds to a unique pseudo-random string associated with the voter. Its knowledge provides no additional information to the $EA$, since the $EA$ knows voters’ identities.

The AoV framework permits participants to change their vote at any time.
The effect of the change is manifested at the end of the epoch when the tally is computed.  
However, network adversary $Adv_{net}$ can observe the vote transactions on $SC$ even though the vote choice remains confidential (preserving the privacy of votes).
A participant might vote in one interval and then vote again in a future interval.
$Adv_{net}$ cannot distinguish whether the participant voted again for the same candidate or changed her vote to a different
candidate due to assumed confidentiality-preserving properties of plugged voting protocols (see \autoref{sec:underlying-voting-protocol})
However, both votes may be mapped to the same participant's blockchain wallet address if utilized naively, hence $Adv_{net}$ can determine how many times a participant voted.

To break the map between participant $P$ and her  wallet address, we modify the idea presented in type 2 deterministic wallets\footnote{See \url{https://en.bitcoin.it/wiki/BIP_0032}.}.
The objective is to synchronize a practically unlimited number of wallet public keys PKs (one per vote) between $EA$ and $P$ such that this PK list can be regenerated only by these two parties, while the corresponding private keys SKs can be computed only by $P$.
As mentioned in \autoref{ssec:advmodel}, $EA$ is assumed to verify identities honestly, and it supplies their corresponding wallet addresses to $SC$.
The wallet address is generated as a function $f$ of the elliptic curve public key. 
Once the public key is available, it is straightforward to compute the corresponding wallet address.
Let $BP$ be the base point on the elliptic curve.
Further, let $PK$ be the blockchain wallet public key corresponding to a private key $SK$.
Here, $SK$ is chosen as a random positive integer whose size is bounded to the order of $BP$ on the chosen elliptic curve modulo a prime number.
Note that \autoref{eqn_anon01} -- \autoref{eqn_anon05}  are computed off-chain.
As an illustration, let the first PK be computed as
\begin{equation}\label{eqn_anon01}
	PK_{0} = SK_{0}\cdot BP,
\end{equation}
and the next PK be
\begin{equation}\label{eqn_anon02}
	PK_{1} = PK_{0}+SK_{1}\cdot BP.
\end{equation}
From \autoref{eqn_anon01} and~\autoref{eqn_anon02}, we observe that
\begin{equation}\label{eqn_anon03}
	(SK_{0}+SK_{1})\cdot BP = PK_{1}.
\end{equation}
The following steps ensue:
\begin{compactenum}[i)]
	\item During the identity verification, $P$ sends to $EA$:
	\textit{a}) wallet public key $PK_{0}$, \textit{b}) a random shared secret key $hk$, and \textit{c}) parameters $(g,p)$, where $g$ is a randomly chosen generator in $F_p^{*}$ (i.e., a prime field)  and $p$ is a large prime.
	The wallet address is a public function of the wallet public key.
	Hence, $EA$ computes $W_{0}=f(PK_0)$ and stores it.
	
	\item The private key of $P$ at any future voting epoch $e=\{1,2,3,...,2^{128}-1\}$  is generated by $P$ as   
	\begin{equation}\label{eqn_anon04}
		SK_{e}= SK_{0} + HMAC_{hk}(g^e),
	\end{equation}
	where $g^e\in F_p^{*}$ is the output of pseudo-random number generator (PRNG) in epoch $e$, HMAC$(.)$ is HMAC-X-256 using shared secret key ${hk}$ between $EA$ and $P$, which is unknown to $Adv_{net}$ and serves for stopping her from mapping $P$'s wallet addresses.
	
	\item The corresponding $PK$ of $P$ for epoch $e$ is   
	
	\begin{equation}\label{eqn_anon05}
		PK_{e}= PK_{0} + HMAC_{hk} (g^e)\cdot BP.
	\end{equation}
\end{compactenum}
$EA$ and $P$ can compute $PK_e$ but $SK_e$ is held only by $P$.
This effectively separates PKs from their SKs and, at the same time, maps it to $P$'s first wallet public key, i.e., $PK_{0}$.   
At any voting iteration $e$, the public key $PK_{e}$, and the corresponding wallet address can be computed by both $EA$ and $P$.
Since the shared secret $hk$ used with HMAC is known only to $EA$ and $P$, no third party, including $Adv_{net}$, is able to compute any future PKs.
Hence, for a sequence of wallet addresses of $P$  given by $W_{e}$ = $f(PK_{e})$, the map between the wallet address and $P$ is broken for all other parties other than $EA$ and $P$.

\subsubsection*{\textbf{Batching the Requests}}
It is important to note that if a participant wishes to change her vote within the same voting interval, she should submit the request with the new wallet address to $EA$ who will approve the request in batches (aggregating multiple such requests) in order to improve the resistance against mapping of former wallet addresses to new ones.
In detail, the $EA$ will mark all former addresses as \textit{invalid} and approve the new ones within a single transaction.
In the extreme case, when a batch contains only one participant (who wanted to change her vote), $Adv_{net}$ can map the two wallets.
Therefore, such a participant should value her wallet privacy and vote only once during the next interval (or alternatively change her address again and be a part of a bigger batch).
To summarize, the privacy of participants is achieved by synchronizing their  wallet addresses with the election authority, so no third-party observer can map any future participants' wallet addresses to the current or previous ones.
Using VPN/dVPNs may further limit the $Adv_{net}$'s ability to map participant IP addresses.

\section{Functionality}
\label{ssec:functionality}

\begin{algorithm}[t]
	\label{alg:vdfdeposit}
	\DontPrintSemicolon
	
	\footnotesize
	\SetKwProg{Fn}{Function}{:}{\KwRet}
	\SetKwFunction{VDFADD}{VdfAdd}
	
	\SetKwProg{Fn}{Def}{:}{}
	\Fn{\VDFADD{$y, \pi, blockheight$}}{    
		writeState($``vdfadd"~||~ blockheight, y~||~\pi$)    
	}
	\caption{VDF Add}
\end{algorithm}

\begin{algorithm}[t]
	\label{alg:bpodeposit}
	\DontPrintSemicolon
	\footnotesize
	
	\SetKwProg{Fn}{Function}{:}{\KwRet}
	\SetKwFunction{BPOAdd}{BpoAdd}
	
	\SetKwProg{Fn}{Def}{:}{}
	\Fn{\BPOAdd{$Target, BH, blockheight$}}{    
		writeState($``bpoadd"~||~ blockheight,BH~||~Target$)\\    
		writeState($``blockheightStored",blockheight$)\\
		writeState($``blockheader"~||~blockheight,BH$)
	}
	\caption{BPO Add}
\end{algorithm}

The high-level functionality of the AoV framework and its smart contracts is shown in Algorithm~\autoref{alg:framework}, and the trigger mechanism is presented in Algorithm~\autoref{alg:trigger}.
Algorithm~\ref{alg:framework} comprises 5 main functions --- \textit{setup, registration, voting, tally} and \textit{revote}.
The system parameters agreed upon are added by $EA$ using the \textit{setup} function of the smart contract.
The $EA$ is also responsible for adding the list of valid participants' wallet addresses to the contract through the  \textit{registration} function.
The \textit{voting} function is supplied by a participant's wallet address, blinded vote, and its proof of correctness.
This information is signed with $P_i$'s private key and sent to the contract.
The \textit{voting} function carries out the necessary verifications and adds her vote. The participant wallet address is set to ``voted'' to disallow its reuse.
Before invoking the \textit{tally} function, the VDF prover and BPO store their respective data to the contract (see Algorithm~\ref{alg:vdfdeposit} and Algorithm~\ref{alg:bpodeposit}).
Next, the \textit{tally} function is called by $EA$ or any authorized participant.  
The \textit{tally} function carries out two main tasks. First, it checks whether the condition to trigger the interval tally is satisfied (see Algorithm~\ref{alg:trigger}). The second task (when triggered) is to tally the votes and return the results.
When a participant wishes to vote again, she sends her next wallet address (synchronized with $EA$) to the \textit{revote} function.
The $EA$ will verify the new wallet address offline and call the \textit{registration} function to set the new address to valid (preferably in batches as mentioned in \autoref{ssec:appendixWallet}).
Next, a participant may call the function \textit{Voting} and vote using her new wallet address.

\begin{algorithm}[t]
	\label{alg:trigger}
	
	\DontPrintSemicolon
	\footnotesize
	
	\SetKwProg{Fn}{Function}{:}{\KwRet}
	\SetKwFunction{FTrigger}{VerifyTrigger}
	
	\SetKwProg{Fn}{Def}{:}{}
	\Fn{\FTrigger{$y, \pi, T, BH, params\_struct$}}{    
		$b$ = -$1$\\
		$tt$ = readState($params\_struct.totaltime$)\\
		$ft$ = readState($params\_struct.ft$)\\
		$seed$ = readState($params\_struct.key$)\\  
		$IntervalTime$ = $tt/ft$\\
		$BHsInInterval$ = $IntervalTime/10$\\
		
		\If{$SHA256(BH)<T$ }
		{
			\If{ $Verify\_VDF(y,\pi)==True$}
			{
				$a = SHA(y)$\;
				$b = a (mod\; BHsInInterval)$\\   		 
			}
		}
		\If{b==0}
		{
			\KwRet $True$\;
		}
		\Else{\KwRet $False$\;}
	}

	\caption{Trigger Mechanism}
\end{algorithm}

\begin{algorithm*}
	\label{alg:framework}
	
	\footnotesize
	\SetKwFunction{FMain}{Main}
	\SetKwFunction{FSetup}{Setup}
	\SetKwFunction{FReg}{Registration}
	\SetKwFunction{FVoting}{Voting}
	\SetKwFunction{FTally}{Tally}
	\SetKwFunction{FRevote}{Revote}
	
	\KwInput{Set1: $\forall$ participants $P_i$, wallet addresses $(WA_{ij})$, blinded vote $BV_{ij}$ (by $P_i$ for her $j^{th}$ voting occurrence,  $j=0,1,2,3,...$) \& zero knowledge proof of vote correctness ($ZKP_{ij}$), $booth_{no}$. Set2: system parameters $init\_params$, BTC blockheader $BH$, $VDF(BH)$, proof $\pi$, BTC target $T$, BTC blockheight.}
	\KwOutput{Total tally of votes in the interval.}
	
	\medskip
	\SetKwProg{Fn}{Function}{:}{}
	\Fn{\FSetup{$init\_params$}}
	{
		writeState($params\_struct, init\_params$) \tcp{add system parameters as key-value pairs into $params\_struct$.}
	}
	\smallskip
	
	\SetKwProg{Fn}{Function}{:}{}
	\Fn{\FReg{$msg1=WA_{ij}, msg2=valid\_flag, EA\_signed\_msg$}}{
		$msg = msg1~||~msg2$ \tcp{concatenate message parts.}
		$EA\_{pubkey}$= readState($params\_struct.EA\_public\_key$) \tcp{get $EA$ public key.}
		
		\If{VerifySig$(msg,EA\_signed\_msg, EA\_{pubkey})=True$}
		{
			\If{$valid\_{flag}==True$}
			{
				writeState($WA_{ij},``valid"$) \tcp{set wallet address to valid.}
			}
			\Else
			{
				writeState($WA_{ij},``invalid"$) \tcp{set wallet address to invalid.}
				
			}
		}
	}
	\smallskip

	\SetKwProg{Fn}{Function}{:}{}
	\Fn{\FVoting{$msg1=WA_{ij},msg2=BV_{ij},msg3=ZKP_{ij},P_i\_signed\_msg$}}{
		$msg = msg1~||~msg2~||~msg3$\\
		$wallet\_status$= readState($WA_{ij}$)\\
		$P_i\_{pubkey}$= readState($params\_struct.P_i\_public\_key$)\\
		$sig\_flag$ = VerifySig($msg$,$P_i\_signed\_msg$,$P_i\_{pubkey}$)\\
		$zkp\_flag$ = VerifyZKP($BV_{ij},ZKP_{ij}$)\\
		
		\If{($sig\_flag$ and $zkp\_flag$) == $True$ and $wallet\_status==``valid"$ }
		{
			writeState($``vote"~||~WA_{ij},BV_{ij}$) \tcp{The latest wallet address of $P_i$ is mapped to her private vote. The key  in (key,value) is prefixed with `vote' tag to identify valid votes w.r.t. wallet addresses.}
			writeState($WA_{ij},``voted"$) \tcp{set $WA$ to voted \& prevent voting from that address again.}

		}
	}
	\smallskip

	\SetKwProg{Fn}{Function}{:}{}
	\Fn{\FTally{$blockheight$}}{
		$total\_tally=-1$\\
		$stored\_blockheight$ = readState($``blockheightStored"$)\tcp{\textit{blockheightStored} is from Algorithm~\ref{alg:bpodeposit}: BPO Add.}
		$BH=$readState($``blockheader"~||~blockheight$)\tcp{\textit{blockheight} is the argument passed to the function Tally.}
		$y,\pi$ = readState($``vdfadd"~||~blockheight$)\tcp{\textit{vdfadd} is read from Algorithm~\ref{alg:vdfdeposit}.}
		$BH,Target$ = readState($``bpoadd"~||~blockheight$)\\
		$trigger\_flag$=VerifyTrigger($y, \pi, Target,BH, params\_struct$)\tcp{ Call Algorithm~\autoref{alg:trigger}.}
		\If{($stored\_blockheight$ == $blockheight$) and $trigger\_flag$ == True}
		{
			
			$total\_tally = \sum\limits_{no=1}^{X\cdot Y} local\_tally(booth_{no})$\\
			
		}
		\KwRet $total\_{tally}$   	 
	}
	\smallskip
	
	\SetKwProg{Fn}{Function}{:}{}
	\Fn{\FRevote{$msg=WA_{ij}, P_i\_signed\_msg$}}{
		$P_i\_{pubkey}$= readState($params\_struct.P_i\_public\_key$) \tcp{get $P_i$ public key}
		
		\If{VerifySig$(msg,P_i\_signed\_msg, P_i\_{pubkey})=True$}
		{
			writeState($WA_{ij},``pending"$) \tcp{set wallet address to pending verification by $EA$.}
		}
		\tcp{Next, $EA$  calls $Registration()$, where it sets a new wallet address of $P_i$ to valid.}
		\tcp{Further, $P_i$ calls $Voting()$ to re-vote using the new wallet address.}
		
	}

	\caption{Always on Voting Framework}
\end{algorithm*}

\section{Privacy Implications of Booth Sharding}
\label{ssec:appendixSharding}

In this section, we look at the implications of booth sharding.
The main reason for using booth sharding is to reduce smart contract invocation costs (considering the voting is implemented on a blockchain), when it involves data processing or on-chain storage. Splitting the booth contract into  a number of smaller contract modules is aimed at reducing the costs paid by its voting participants and election authority  towards smart contract computations.

Further, we recommend booth sizes to protect participant votes from being revealed to the $EA$ (in the case when all participants in a booth voted for the same candidate).
The $EA$ is aware of the participants' current wallet address used for voting, and hence able to make statistical inferences.
Certain scenarios revealing vote choice are possible under some circumstances.
In particular, if all participants in a booth voted for the same candidate or the winning probability of one candidate is much higher than the others.

To demonstrate it, in \autoref{fig:binomialpdf} we provide the probability that $X$ participants in a booth voted for the same candidate, depending on the candidate winning probability $p$.
\autoref{fig:binomialpdf}(a) represents a booth with 30 participants and candidate winning probability $p=0.9$. The probability that all participants voted for the same candidate is
$P(30, X=30,p=0.9)\stackrel{\sim}{=} 0.0423$.
\autoref{fig:binomialpdf}(b) represents a booth with  100 participants and candidate winning probability $p=0.9$. The probability that all participants voted for the same candidate is
$P(100, X = 100,p=0.9)\stackrel{\sim}{=} 0.00003$, which demonstrates that the number of participants in a booth influences $p$ in indirect proportion, favoring the booths with higher sizes.

Even though the probabilities in the booth with 100 participants are very low, this may not be sufficient depending on the total number of participants.
For example, consider the elections with 1 million participants.
First, let the number of participants in a booth be 30 and the number of booths $M$ $ = \lceil(1~000~000/30)\rceil =33334$.
The number of booths where all participants likely voted for the same candidate is $0.0423\cdot33334 \stackrel{\sim}{=} 1410$.
For booths with 100 participants each and $M = \lceil(1~000~000/100)\rceil = 10000$, the number of booths where all participants likely voted for the same candidate is reduced to $0.00003\cdot10000 \stackrel{\sim}{=} 0.3$. Therefore, a suitable number of participants per booth should be determined based on the extreme estimations of the tally results and the total number of voters.




\ifCLASSOPTIONcaptionsoff
  \newpage
\fi

\end{document}